\documentclass[aoas,preprint]{imsart}

\RequirePackage{amsthm,amsmath,amsfonts,amssymb}
\RequirePackage[authoryear]{natbib}
\RequirePackage[colorlinks,citecolor=blue,urlcolor=blue]{hyperref}
\RequirePackage{graphicx}

\usepackage{color}
\usepackage{wasysym}
\usepackage{float}
\usepackage{bm}
\usepackage{bbm}
\usepackage{mathtools}
\usepackage{verbatim}  
\usepackage{enumerate}
\usepackage{fancyvrb}
\usepackage{dsfont}
\usepackage{changepage}
\usepackage{multirow}
\usepackage{pdflscape}
\usepackage{comment}
\usepackage[figuresleft]{rotating}
\usepackage{xparse}

\startlocaldefs
\endlocaldefs

\newcommand{\expit}{\mbox{expit}}

\NewDocumentCommand{\evalat}{sO{\big}mm}{%
  \IfBooleanTF{#1}
   {\mleft. #3 \mright|_{#4}}
   {#3#2|_{#4}}%
}

\begin{document}

\begin{frontmatter}

\title{Space-time smoothing models for sub-national measles routine immunization coverage estimation with complex survey data\thanksref{t1}}
\runtitle{Space-time smoothing for RI-specific MCV1 coverage}
\thankstext{T1}{Supported by NIH Grant R01-AI029168.}

\begin{aug}
\author[A]{\fnms{Tracy Qi} \snm{Dong}\ead[label=e1, mark]{qd8@uw.edu}}
\and
\author[A]{\fnms{Jon} \snm{Wakefield}}

\address[A]{University of Washington \printead{e1}}
\end{aug}

\begin{abstract}
Despite substantial advances in global measles vaccination, measles disease burden remains high in many low- and middle-income countries. A key public health strategy for controling measles in such high-burden settings is to conduct supplementary immunization activities (SIAs) in the form of mass vaccination campaigns, in addition to delivering scheduled vaccination through routine immunization (RI) programs. To achieve balanced implementations of RI and SIAs, robust measurement of sub-national RI-specific coverage is crucial. In this paper, we develop a space-time smoothing model for estimating RI-specific coverage of the first dose of measles-containing-vaccines (MCV1) at sub-national level using complex survey data. The application that motivated this work is estimation of the RI-specific MCV1 coverage in Nigeria's 36 states and the Federal Capital Territory. Data come from four Demographic and Health Surveys, three Multiple Indicator Cluster Surveys, and two National Nutrition and Health Surveys conducted in Nigeria between 2003 and 2018. Our method incorporates information from the SIA calendar published by the World Health Organization and accounts for the impact of SIAs on the overall MCV1 coverage, as measured by cross-sectional surveys. The model can be used to analyze data from multiple surveys with different data collection schemes and construct coverage estimates with uncertainty that reflects the various sampling designs. Implementation of our method can be done efficiently using integrated nested Laplace approximation (INLA). 
\end{abstract}


\end{frontmatter}


\section{Introduction} \label{Chap3SecIntro}

Measles is a highly contagious, vaccine-preventable disease. Despite tremendous advances in the global measles vaccination coverage in the past few decades, challenges remain in many low- and middle-income countries (LMICs), where disease burdens are the highest \citep{IA2030}. A key public health strategy in such high-burden settings is the combination of routine immunization (RI) and supplementary immunization activities (SIAs) \citep{mri}. Although the World Health Organization (WHO) recommends two doses of measles-containing-vaccine (MCV) for all children --- first at the age of 9 months and second at the age of 15 months, most LMICs only administer one dose in their RI schedule \citep{who_measles1}. In order to increase measles vaccination coverage, SIAs are carried out in the form of mass immunization campaigns every 2--4 years. During these campaigns, health workers run fixed-post vaccination sites and provide MCV to all children in a pre-specified target age group, regardless of whether they have been vaccinated previously. For large countries where there is substantial heterogeneity in disease burden and RI coverage across geographical regions, SIA campaigns are often designed and implemented based on each region's specific situations. For example, the SIA campaigns in Nigeria's 20 northern states have been implemented at a different schedule, and sometimes with a different target age group, compared to the SIAs campaigns in its 17 southern states, as shown in Table S2 of the Supplementary Material \citep{dong2020supp}.

To achieve balanced implementations of RI and SIAs against measles, robust measurement of sub-national RI-specific coverage is crucial. Program officers rely on accurate estimation of local RI coverage to identify areas where the routine vaccine delivery systems are the weakest \citep{biellik2018strengthening, mri1} or the SIA campaigns are the most effective \citep{utazi2020geospatial}. In addition, RI-specific coverage of the first dose of MCV (MCV1) is an essential input to measles epidemiological models for estimating underlying susceptible population dynamics and informing optimal SIA schedules \citep{verguet2015controlling, thakkar2019decreasing}. Specifically, in order to calculate the rate at which the susceptible population grows over time, one needs to estimate the number of infants who are not protected by vaccination after losing their maternal immunity against measles at the age of approximately 9 months. This requires reliable estimation of the RI-specific MCV1 coverage across different birth cohorts and sub-national areas. 

Despite ongoing efforts to strengthen administrative data quality \citep{bid}, the health management information systems in many LMICs are too weak to provide data of adequate quality for assessing and guiding health programs \citep{hancioglu2013measuring}. Therefore, household surveys are the primary data sources for vaccination coverage estimates \citep{cutts2016monitoring, dolan2017comparison}. The Demographic and Health Surveys (DHS) \citep{DHS} and the Multiple Indicator Cluster Surveys (MICS) \citep{MICS} are two major survey programs that report sub-national measles vaccination coverage estimates. In addition, geospatial modeling approaches have been applied to the household survey data to generate MCV coverage estimates at various geographical scales \citep{utazi2018high, utazi2018spatial, utazi2019mapping}. However, all the above-mentioned estimates are not RI-specific --- they reflect the overall coverage of measles vaccination received via RI, SIAs or both sources. These coverage estimates can be influenced by past vaccination campaigns, and hence are not suitable for the aforementioned RI-specific purposes. For example, using the overall MCV coverage in epidemiological models may lead to under-estimation of the rate at which the susceptible population grows, resulting in inaccurate estimation of measles transmission dynamics. To obtain RI-specific coverage estimates, the current method relies on only using data from birth cohorts that have not had any SIA opportunities yet. This restriction may result in some or all data from a household survey not being eligible for analysis --- an inefficient use of already sparse data. 

In this paper, we propose a space-time smoothing model for estimating sub-national RI-specific MCV1 coverage using data from complex surveys. We demonstrate the method by calculating the state-level RI-specific MCV1 coverage estimates in Nigeria using data from 8 household surveys conducted between 2003 and 2018. Inspired by the small area estimation methods developed by \cite{mercer2015space}, our model is able to combine data from multiple surveys with different sampling designs and construct sub-national coverage estimates with uncertainty that reflects the various data collection schemes. Our method allows more efficient use of data by modeling survey outcome not just from children who only had RI, but also from those who had RI and an SIA opportunity. We account for the impact of SIAs by incorporating information from the WHO SIA calendar \citep{who_measles3}. 

This paper is structured as follows. In Section \ref{Chap3SecData}, we describe the Nigeria data upon which estimation will be based. The details of our proposed method are provided in Section \ref{Chap3SecMethod}. The results of our modeling efforts of RI-specific MCV1 coverage in Nigeria from 2003 to 2018 are presented in Section \ref{Chap3SecApp}. Section \ref{Chap3SecDisc} contains concluding remarks. 


\section{Data sources} \label{Chap3SecData}

We focus on the state-level RI-specific MCV1 coverage in Nigeria using data from four DHS surveys conducted in 2003, 2008, 2013 and 2018, three MICS surveys conducted in 2007, 2011 and 2016--17\footnote{The 2016--17 MICS is combined with the National Immunization Coverage Survey (NICS).}, and two National Nutrition and Health Surveys (NNHS) conducted in 2014 and 2015. In addition, we obtain the SIA calendar from the WHO website, which contains the start and end dates, target age groups and geographic areas of all SIAs implemented in Nigeria. More details about the data are provided below. 


\subsection{Household surveys}

The DHS, MICS and NNHS used similar, but not identical stratified cluster sampling designs to provide vaccination coverage estimates at the national level and in each of Nigeria's 36 states and the Federal Capital Territory (FCT) \citep{dong2020impact}. Hereafter we refer to these as \textit{Nigeria's 37 states}. The DHS surveys were stratified by urban and rural areas within each state, whereas MICS and NNHS were stratified only by state. Selection of survey respondents took place in three stages: First, within each stratum, primary sampling units (PSUs or \textit{clusters}) were selected from a census frame of enumeration areas (EAs). The 2003 DHS and 2007 MICS used frames based on the 1991 census and the other surveys used frames based on the 2006 census. Second, within each selected cluster, a pre-specified number of households were randomly selected from an updated listing of all residential households. Third, within each selected household, field teams attempted to interview all women aged 15--49 years and collect vaccination data for children under age 3 or 5 years. 

Every child in the survey is associated with a survey weight that quantifies the relative number of children he or she represents in the population. Generally, DHS and MICS surveys calculate a base weight for each child as the inverse of the product of the probabilities of selection from each sampling stage. The base weights are then adjusted to reflect household or individual response rates. In contrast, the NNHS were conducted based on the Standardized Monitoring and Assessment of Relief and Transitions (SMART) method \citep{SMARTv1, SMARTv2}: each child was assigned the same base weight across all clusters, and the base weights were then post-stratified to match a set of state population proportions without further adjustment for the non-response rate. Table S1 of the Supplementary Material summarizes key information regarding the sampling design of each survey, including source of the sampling frame, survey strata, cluster and household sample sizes, age group with vaccination data, and features of weight calculation. 

For each child sampled in a survey, the care giver was asked to provide information about specific vaccines that the child had received up until the time of interview. Specifically, care givers were first asked to present any home-based record (HBR), such as vaccination cards, to the survey interviewer. If HBR is available, the interviewer would populate a designated section of the survey questionnaire that detail the types and doses of vaccines received by the child. After this step is complete, the interviewer would ask the care giver a follow-up question of whether additional vaccine doses had been administered besides the ones listed on the HBR. This would include vaccines received at health care facilities via RI but which had not been recorded on paper, or those received via SIAs. If additional doses were administered, the interviewer will go back to the same section and record any additional doses based on the care giver's recall. If no HBR is available, the interviewer would skip the HBR section and ask the care giver to recall whether the child has received each specific vaccine at the time of survey interview, regardless of whether the vaccination is administered through RI or SIAs. As such, the surveys were designed to measure the percentage of children in various birth cohorts who received specific vaccines at any time before the survey through any source, and it is extremely difficult, if not impossible, to differentiate whether a child received MCV1 through RI or SIAs based on the answers to the survey questionnaires alone. 


\subsection{SIA calendar}

The WHO SIA calendar is a publicly available data source where key information on the completed and planned SIAs in member states is provided \citep{who_measles3}. For each SIA, the calendar records the start and end dates, target age group, size of the target population and geographical areas in which the campaign is carried out. Table S2 of the Supplementary Material  shows the details of the SIAs in Nigeria conducted before the end of 2018 that we consider for this analysis. Excluded from the analysis are two mop-up campaigns implemented in 2007 and one outbreak response campaign implemented in 2013. These campaigns' scales were very small in terms of target population and geographic area, and therefore were assumed to have negligible impact on the state-level MCV1 coverage. 


\section{Methods} \label{Chap3SecMethod}

The basic idea of our method is that we consider the RI-specific MCV1 coverage as the percentage of children in a birth cohort who received a first dose of measles vaccine through the routine vaccine delivery system according to the RI schedule, i.e., at the age of 9 months. Each cross-sectional survey serves as a snap-shot measurement of the MCV1 coverage within a birth cohort at the time of survey interview. We can carefully align the birth time, RI schedule, SIA schedule and survey time for each child to determine whether a birth cohort has had the RI opportunity or any additional SIA opportunity when the survey data are collected. 

\begin{figure}
    \centering
	 \includegraphics[width=\textwidth]{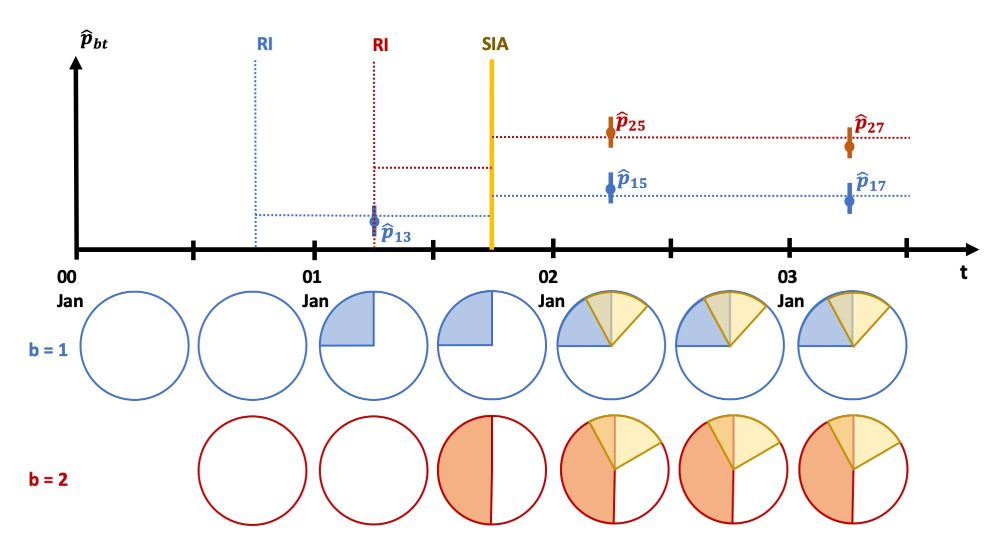}
	\caption{Illustration of the data generating mechanism for a generic area. We let $b$ index 6-month birth cohort and consider two birth cohorts --- cohort $b = 1$ represented by the blue circles, and cohort $b = 2$ represented by the orange circles. Each circle shows the underlying MCV1 coverage of the birth cohort at a specific time. The blue/orange shade represents the proportion of children covered by RI, i.e., the RI-specific MCV1 coverage; and the yellow shade represents the proportion covered by SIA. Because all children within the target age group can receive vaccination during an SIA, regardless of whether they have been vaccinated before, some children in the birth cohorts might be covered by both RI and SIA. They are represented by the intersection of the blue/orange shade and the yellow shade. }
	\label{Chap3FigRI}
\end{figure}

Figure \ref{Chap3FigRI} illustrates the data generating mechanism we assume for a generic area. We discretize time into 6-month intervals and let the beginning of 2000 be the reference starting point. We let $b$ index birth cohort, such that the children born between Jan. and Jun. in 2000 belong to birth cohort $b = 1$, and the children born between Jul. and Dec. in 2000 belong to birth cohort $b = 2$, and so on. We let $t$ index survey time, such that a survey conducted between Jan. and Jun. in 2001 has $t = 3$, and a survey conducted between Jan. and Jun. in 2002 has $t = 5$, and so on. We assume each child is scheduled to receive MCV1 through RI at the age of 9 months. 

As an example, suppose an SIA targeting children aged 9--59 months is implemented between Jun. and Dec. 2001, and consider two birth cohorts in detail --- cohort $b = 1$ represented by the blue circles, and cohort $b = 2$ represented by the orange circles. Each circle indicates the underlying MCV1 coverage of the birth cohort at a specific time. The blue/orange shade represents the proportion of children covered by RI, i.e., the RI-specific MCV1 coverage; and the yellow shade represent the proportion covered by SIA. Because all children within the target age group can receive vaccination during an SIA regardless of whether they have been vaccinated before, some children in the birth cohorts might be covered by both RI and SIA. They are represented by the intersection of the blue/orange shade and the yellow shade. 

Now we consider three annual cross-sectional surveys conducted at time $t = 3, 5 \mbox{ and } 7$ respectively. Let $p_{b,t}$ denote the proportion of children in birth cohort $b$ who have received MCV1 by time $t$. At $t = 3$, children in cohort $b = 1$ should already have had an RI opportunity to receive MCV1, but those in cohort $b = 2$ are still too young to be vaccinated against measles. Therefore, we can use the survey data to obtain a design-based weighted estimate of the MCV1 coverage for birth cohort $b = 1$ at survey time $t = 3$, which we denote $\hat{p}_{1,3}$. In Figure \ref{Chap3FigRI}, $\hat{p}_{1,3}$ and its associated uncertainty interval are represented by a blue dot with a vertical line segment. Because the birth cohort has not been impacted by any SIA at the time of survey, $\hat{p}_{1,3}$ gives an estimate of the RI-specific MCV1 coverage. In the next survey at $t = 5$, both birth cohorts should have had their RI opportunities. In addition, both birth cohorts should also have been targeted during the SIA implemented in the previous time interval. Since survey data cannot differentiate the source of the vaccination, the design-based estimates of the MCV1 coverage for birth cohort $b = 1 \mbox{ and } 2$ at time $t = 5$, denoted $\hat{p}_{1,5}$ and $\hat{p}_{2,5}$, would reflect the overall MCV1 coverage in these birth cohorts instead of the RI-specific coverage. In Figure \ref{Chap3FigRI}, the overall MCV1 coverage is represented by the union of the blue/orange shade and the yellow shade. The same logic applies to the survey conducted at $t = 7$. 

As mentioned in Section \ref{Chap3SecIntro}, our method uses survey data from children who only had RI and children who had RI and one SIA opportunity. Based on the data generating mechanism described above, our proposed approach takes three steps to construct sub-national RI-specific MCV1 coverage over time: first, individual-level data from each survey is processed to identify eligible birth cohorts; second, design-based MCV1 coverage estimates and associated variances are calculated for all eligible birth cohorts identified in step one at the sub-national area level; last, a space-time smoothing model is applied to the design-based estimates from step two to predict RI-specific MCV1 coverage over time for all sub-national areas. We now describe each step in detail.


\subsection{Data processing} \label{Chap3SecDataproc}

In each survey, the time of interview for each child is recorded in Century-Month Code (CMC) format. In addition, the child's age in months at the time of interview is also recorded. Combining the two pieces of information allows us to identify the birth month, and hence the birth cohort, for each child. 

Assuming every child has their RI opportunity at the age of 9 months, we use the age information of each child to see whether he/she has had the RI opportunity at the time of the survey. We also align the SIA calendar with each child's birth month to identify how many SIA opportunities he/she has had before the survey interview. 

With each child's birth cohort, RI and SIA information collected, we now identify, within each sub-national area, the birth cohorts in which all children only had their RI opportunity, or all children had RI plus one SIA opportunity in each survey. Data from these eligible birth cohorts will be used in the next two steps. 


\subsection{Design-based overall MCV1 coverage estimation} \label{Chap3SecDirest}

For each eligible birth cohort identified in each survey, the design-based overall MCV1 coverage estimate is calculated at the sub-national area level. Let $i$, $b$, $s$ and $k$ be the indices for area, birth cohort, survey and sampled child respectively. We obtain the Horvitz-Thompson (HT) direct estimate \citep{horvitz1952generalization} of the overall MCV1 coverage in birth cohort $b$ in area $i$ from survey $s$, denoted $\hat{p}_{ibs}$, using
\begin{gather*}
    \hat{p}_{ibs} = \frac{\sum_{k = 1}^{n_{ibs}} w_{ibsk} y_{ibsk}}{\sum_{k = 1}^{n_{ibs}} w_{ibsk}},
\end{gather*}
where $y_{ibsk}$ is the 0/1 indicator of whether child $k$ in birth cohort $b$ in area $i$ has ever received MCV1 when interviewed for survey $s$, $w_{ibsk}$ is the survey weight associated to the child, and $n_{ibs}$ is the total number of children in birth cohort $b$ in area $i$ who are sampled in survey $s$. In addition, we also obtain the design-based estimate of the variance associated with $\hat{p}_{ibs}$, which we denote $\hat{v}_{ibs}$. All design-based estimates can be computed using functions from the \texttt{survey} package \citep{lumley2004analysis} within the \texttt{R} computing environment \citep{R2020}.


\subsection{Space-time smoothing model for RI-specific MCV1 coverage estimation} \label{Chap3SecSPmod}

We extend the Fay-Herriot approach \citep{fay1979estimates} developed in \cite{mercer2015space} and specify a Bayesian hierarchical space-time smoothing model to construct sub-national RI-specific MCV1 coverage estimates over time with uncertainty that reflects the various survey sampling designs.

First, the design-based estimates for the logit-transformed overall coverage and its associated variance, denoted by $\hat{\theta}_{ibs}$ and $\hat{V}_{ibs}$, are obtained via
\begin{gather*}
    \hat{\theta}_{ibs} = \log \left(\frac{\hat{p}_{ibs}}{1 - \hat{p}_{ibs}} \right), \\
    \hat{V}_{ibs} =  \left(\frac{\hat{v}_{ibs}}{\hat{p}_{ibs}^2 (1 - \hat{p}_{ibs}^2)} \right).
\end{gather*} 

Next, we take the following asymptotic distribution as a working likelihood:
\begin{gather}
    \hat{\theta}_{ibs} | \theta_{ibs} \sim \mbox{Normal}\left( \theta_{ibs}, \hat{V}_{ibs} \right) \label{Chap3Eqn1}
\end{gather}
where $\theta_{ibs}$ is the underlying logit-transformed overall coverage in cohort $b$ in area $i$ measured by survey $s$. To account for the potential impact of SIA and specific surveys on the overall coverage, we specify $\theta_{ibs}$ as
\begin{align}
    \theta_{ibs}  (x_{ibs})  &= \mu_{ib} +  \beta_1 \times x_{ibs} + \epsilon_s \label{Chap3Eqn2} \\
    \mu_{ib} &= \beta_0 + \alpha_i + \gamma_i + \delta_b + \tau_b + \phi_{ib} \label{Chap3Eqn3}
\end{align}
The potential impact of SIA on the overall MCV1 coverage is captured by $\beta_1 \times x_{ibs}$, where $x_{ibs}$ is an 0/1 indicator of whether children in birth cohort $b$ in area $i$ has had one SIA opportunity in addition to RI in survey $s$, and $\beta_1$ is a parameter quantifying the impact. We would expect $\beta_1 > 0$. In addition, we use the survey random effects $\epsilon_s \sim \mbox{Normal} \left( 0, \sigma_{\epsilon} \right)$ to capture any survey-specific biases, although this is relative to the average of all the surveys, and does not correct for any overall bias in the surveys combined. 

The underlying RI-specific MCV1 coverage in birth cohort $b$ in area $i$ is captured by the RI term $\mu_{ib}$. Note that this quantity remains unchanged across surveys. Since we consider the RI-specific MCV1 coverage to be the percentage of children in a birth cohort who received a first dose of MCV through the routine immunization system at the age of 9 months, the underlying RI-specific coverage of a birth cohort only depends on when and where the children are born. 

As such, if a birth cohort $b$ in area $i$ only had RI by the time of a survey $s$, $x_{ibs}$ would be zero and the underlying overall coverage $p_{ibs}$ measured in the survey would be
\begin{gather*}
    \evalat{p_{ibs} (x_{ibs})}{x_{ibs} = 0} = \expit \left( \evalat{\theta_{ibs} (x_{ibs})}{x_{ibs} = 0} \right) = \expit \left( \mu_{ib} + \epsilon_s \right).
\end{gather*}
If the birth cohort has had one SIA opportunity in addition to RI by the time of the survey, the underlying overall coverage $p_{ibs}$ measured in the survey would be
\begin{gather*}
    \evalat{p_{ibs} (x_{ibs})}{x_{ibs} = 1} = \expit \left( \evalat{\theta_{ibs} (x_{ibs})}{x_{ibs} = 1} \right) = \expit \left( \mu_{ib} + \beta_1 + \epsilon_s \right).
\end{gather*}
Note that ecological fallacy is not a concern here, because $x_{ibs}$ is constant within areas. Since we only use data from birth cohorts in which all children only had their RI opportunity ($x_{ibs} = 0$), or all children had RI plus one SIA opportunity ($x_{ibs} = 1$) in a survey, there is no individual-level heterogeneity in $x_{ibs}$ within each birth cohort. 

Space-time smoothing is achieved by modeling the RI term $\mu_{ib}$ as a combination of six components, see Equation (\ref{Chap3Eqn3}). Specifically, $\beta_0$ is a common intercept for all birth cohorts and areas. There are two spatial terms that correspond to the convolution model of \cite{besag1991bayesian}. In particular, $\alpha_i \sim \mbox{ICAR}(\sigma_{\alpha})$ are the intrinsic conditional autoregressive (ICAR) terms (described in Section S1.1 of the Supplementary Material) for $i = 1, \dots, I$ sub-national areas (e.g., $I = 37$ for Nigeria's  37 states), and $\gamma_i \sim_{iid} \mbox{Normal}(0, \sigma_{\gamma})$ are independent and identically distributed (IID) random effects. There are also two temporal terms, with $\delta_b \sim \mbox{RW2}(\sigma_{\delta})$ following an (intrinsic) random walk of order 2 model (described in Section S1.2 of the Supplementary Material) to pick up structured temporal trends in RI-specific coverage, and $\tau_b \sim_{iid} \mbox{Normal}(0, \sigma_{\tau})$ being IID to pick up short-term fluctuations with no structure. Finally, the space-time interaction term $\phi_{ib}$ can be specified as one of the four types of interactions described in \cite{knorr2000bayesian} to capture any additional independent or structured variation in RI coverage across space over time. 

To construct RI-specific MCV1 coverage estimates and associated uncertainty intervals for each sub-national area, we use the Bayesian posterior samples of the RI component of the space-time smoothing model. Specifically, the percentage of children in birth cohort $b$ in area $i$ who received a first dose of MCV via RI, $p_{\text{\tiny RI}, ib}$, can be estimated by drawing posterior samples
\begin{gather*}
p_{\text{\tiny RI}, ib}^{(m)} = \expit \left( \mu_{ib}^{(m)} \right) = \expit \left( \beta_0^{(m)} + \alpha_i^{(m)} + \gamma_i^{(m)} + \delta_b^{(m)} + \tau_b^{(m)} + \phi_{ib}^{(m)} \right), \label{Chap3Eqn4}
\end{gather*}
where the superscript $(m)$ denotes the $m$th posterior sample of the respective component. The survey specific effects $\epsilon_{s}$ are not included because they are bias terms. 

Computation in this step, including model fitting and prediction, can be efficiently carried out using the Integrated Nested Laplace Approximation (INLA) method \citep{rue2009approximate} as implemented in the \texttt{INLA} package in \texttt{R}. In Section S4 of the Supplementary Material, we present a simulation study that demonstrates how our approach performs under a number of scenarios with surveys of various sizes. The simulation results also show that the proposed model improves estimation in terms of accuracy and precision through more efficient use of data under each scenario. 


\section{Applying methods to household surveys in Nigeria} \label{Chap3SecApp}

We apply our proposed method to estimate the state-level RI-specific MCV1 coverage in Nigeria between 2000 and 2018 using the household survey and SIA calendar data described in Section \ref{Chap3SecData}. We discretize time into 6-month intervals and let the beginning of 2000 be the reference starting point. During this time period, all children in Nigeria were scheduled to receive one dose of MCV via RI at the target age of 9 months.


\subsection{Data processing and design-based overall MCV1 coverage estimation} \label{Chap3SecApp1}

\begin{figure}
	\centering
	 \includegraphics[width=1\textwidth]{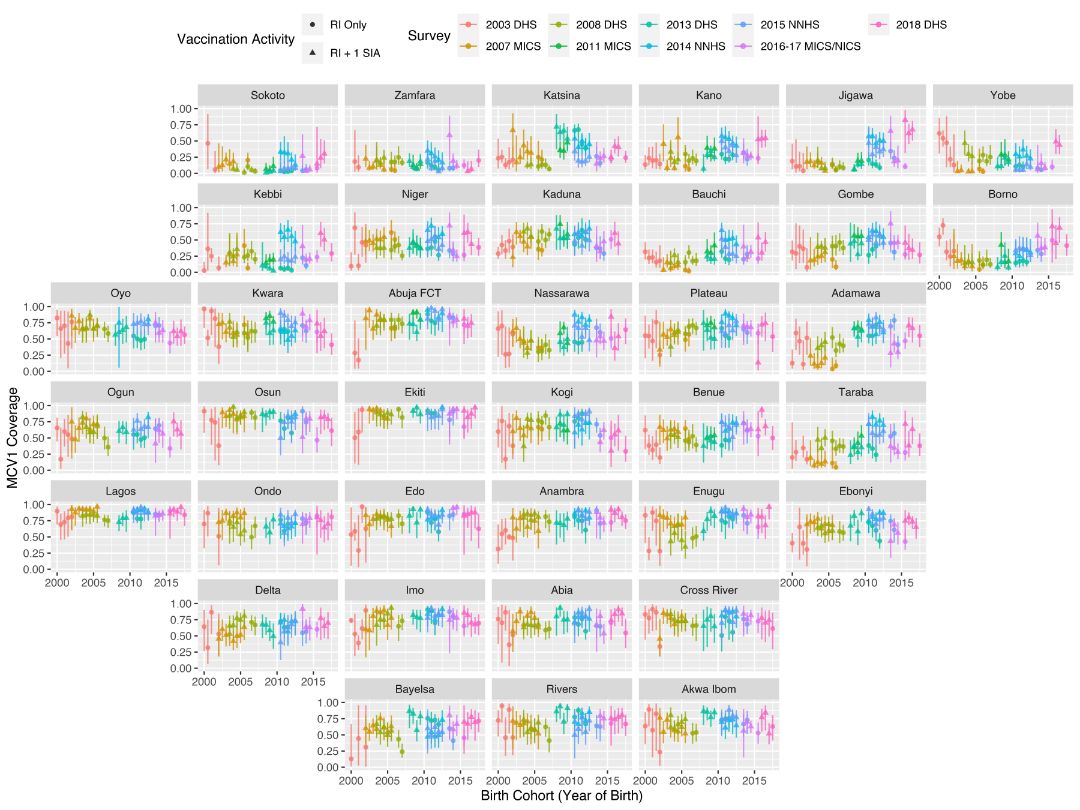}
	\caption{Design-based HT estimates of the overall MCV1 coverage across birth cohorts from each survey for Nigeria's 37 states. }
	\label{Chap3FigDirest}
\end{figure}

We carry out the data processing step outlined in Section \ref{Chap3SecDataproc} to identify, within each state, the birth cohorts in which all children only had their RI opportunity, or all children had RI plus one SIA opportunity in each survey. The design-based HT estimates of the state-level overall MCV1 coverage are calculated for each eligible birth cohort in each survey using the \texttt{survey} package \citep{lumley2004analysis}. Figure \ref{Chap3FigDirest} shows the resultant estimates with the associated confidence intervals across birth cohorts for Nigeria's 37 states. The plots are arranged to approximately match the relative geographical locations of the states. The different colors represent the surveys upon which the design-based estimates are calculated, and the shape of the dots indicates whether a birth cohort had \textit{RI Only} or had \textit{RI + 1 SIA} at the time of survey. The northern states generally have lower MCV1 coverage estimates than the southern states. As expected, within the same survey, the birth cohorts with an additional SIA opportunity tend to have higher estimated MCV1 coverage than the RI-Only cohorts, although the differences are often small in magnitude and difficult to determine here because of the large uncertainty and lack of a modeling framework. The coverage estimates of the same birth cohort from different surveys sometimes show large differences even after accounting for vaccination activity, suggesting that there might be some survey-specific effects. As a result of small sample sizes in some areas and surveys, the design-based coverage estimates for some birth cohorts are associated with very wide confidence intervals. 

\subsection{RI-specific MCV1 coverage estimation via space-time smoothing}

We fit six space-time smoothing models to the design-based estimates. Each model has the same working likelihood and model components as specified in Equations (\ref{Chap3Eqn1}) -- (\ref{Chap3Eqn3}), but a different specification of the space-time interaction term $\phi_{ib}$. Specifically, we consider the \textit{IID-IID}, \textit{ICAR-IID}, \textit{IID-RW2}, \textit{ICAR-RW2}, \textit{IID-AR1}, \textit{ICAR-AR1} models, whose space-time interaction assumptions are detailed in Table \ref{Chap3TabMod1to4} \citep{knorr2000bayesian}. In particular, \textit{AR1} stands for the autoregressive (AR) model of order 1 for Gaussian random vectors. The definition of the AR1 model can found in Section S1.3 of the Supplementary Material.

\begin{table}
\centering
\caption{The space-time interaction assumptions of the models. }
\label{Chap3TabMod1to4}
\begin{tabular}{ll}
\hline
Model & Space-time Interaction $\phi_{ib}$ \\ \hline
\textit{IID-IID} & IID spatial effects  $\times$ IID temporal effects  \\
\textit{ICAR-IID} & ICAR spatial effects  $\times$ IID temporal effects  \\
\textit{IID-RW2} & IID spatial effects  $\times$ RW2 temporal effects  \\
\textit{ICAR-RW2} & ICAR spatial effects  $\times$ RW2 temporal effects  \\
\textit{IID-AR1} & IID spatial effects  $\times$ AR1 temporal effects  \\
\textit{ICAR-AR1} & ICAR spatial effects  $\times$ AR1 temporal effects  \\\hline
\end{tabular}
\end{table}

Implementation of the models was carried out in \texttt{R} using the \texttt{INLA} package. We used the Penalized Complexity (PC) priors of \cite{simpson2017penalising} for all hyper-parameters. Specifically, we applied weak PC priors on the precision parameters for the random effects, such that $\mbox{Pr}(\sigma > 2) = 0.01$ for the spatial and temporal terms, and $\mbox{Pr}(\sigma > 1) = 0.01$ for the independent terms. 

In Table \ref{Chap3TabModSumm}, we report the posterior medians and 95\% CIs of the model parameters. In addition, we also compute three model assessment metrics: the deviance information criterion (DIC), the Watanabe-Akaike information criterion (WAIC) and the log-score conditional predictive ordinate (LCPO) values, all of which are described in detail in Section S2 of the Supplementary Material. Based on the estimates for $\beta_1$, the odds of receiving at least one dose of MCV is 33.6\% -- 37.7\% higher after an SIA campaign is conducted in a birth cohort. The \textit{ICAR-AR1 Model} performs the best among the six models based on all three model metrics. 

\begin{sidewaystable}

\begin{adjustwidth}{1.2cm}{}
\caption{Summary of model assessment metrics and parameter estimates. Bold figures represent the ``best'' models according to the relevant criteria. The posterior medians are taken as the point estimates and the corresponding posterior 95\% CIs are shown in the parentheses. }
\label{Chap3TabModSumm}
\resizebox{1\columnwidth}{!}{%
\begin{tabular}{lcccccccccccc}
\hline
\multirow{3}{*}{Model} & \multicolumn{3}{c}{Model Assessment Metric} & \multicolumn{9}{c}{Parameter Estimate} \\
 & DIC & WAIC & LCPO & Intercept & SIA & ICAR & Area IID & RW2 & Time IID & Space$\times$Time & Survey & AR1 \\ 
 &  &  & & $\beta_0$ & $\beta_1$ & $\sigma_{\alpha}$ & $\sigma_{\gamma}$ & $\sigma_{\delta}$ & $\sigma_{\tau}$ & $\sigma_{\phi}$ & $\sigma_{\epsilon}$ & $\rho_{\phi}$ \\ \hline
IID-IID & 3387 & 3616 & 1.097 & \begin{tabular}[c]{@{}c@{}}-0.013\\ (-0.211, 0.203)\end{tabular} & \begin{tabular}[c]{@{}c@{}}0.32\\ (0.19, 0.43)\end{tabular} & \begin{tabular}[c]{@{}c@{}}0.67\\ (0.53, 0.87)\end{tabular} & \begin{tabular}[c]{@{}c@{}}0.072\\ (0.010, 0.243)\end{tabular} & \begin{tabular}[c]{@{}c@{}}0.25\\ (0.10, 0.57)\end{tabular} & \begin{tabular}[c]{@{}c@{}}0.074\\ (0.038, 0.136)\end{tabular} & \begin{tabular}[c]{@{}c@{}}0.34\\ (0.31, 0.37)\end{tabular} & \begin{tabular}[c]{@{}c@{}}0.23\\ (0.13, 0.45)\end{tabular} & \\ \hline
ICAR-IID & 3344 & 3564 & 1.068 & \begin{tabular}[c]{@{}c@{}}-0.007\\ (-0.208, 0.218)\end{tabular} & \begin{tabular}[c]{@{}c@{}}0.32\\ (0.17, 0.43)\end{tabular} & \begin{tabular}[c]{@{}c@{}}0.67\\ (0.53, 0.87)\end{tabular} & \begin{tabular}[c]{@{}c@{}}0.074\\ (0.010, 0.245)\end{tabular} & \begin{tabular}[c]{@{}c@{}}0.23\\ (0.09, 0.54)\end{tabular} & \begin{tabular}[c]{@{}c@{}}0.091\\ (0.057, 0.139)\end{tabular} & \begin{tabular}[c]{@{}c@{}}0.34\\ (0.31, 0.38)\end{tabular} & \begin{tabular}[c]{@{}c@{}}0.25\\ (0.14, 0.53)\end{tabular} & \\ \hline
IID-RW2 & 3250 & 3386 & 0.984 & \begin{tabular}[c]{@{}c@{}}0.001\\ (-0.203, 0.205)\end{tabular} & \begin{tabular}[c]{@{}c@{}}0.29\\ (0.19, 0.37)\end{tabular} & \begin{tabular}[c]{@{}c@{}}0.25\\ (0.12, 0.46)\end{tabular} & \begin{tabular}[c]{@{}c@{}}0.062\\ (0.009, 0.216)\end{tabular} & \begin{tabular}[c]{@{}c@{}}0.19\\ (0.07, 0.49)\end{tabular} & \begin{tabular}[c]{@{}c@{}}0.089\\ (0.057, 0.136)\end{tabular} & \begin{tabular}[c]{@{}c@{}}0.51\\ (0.40, 0.63)\end{tabular} & \begin{tabular}[c]{@{}c@{}}0.24\\ (0.14, 0.45)\end{tabular} & \\ \hline
ICAR-RW2 & 3206 & 3341 & 0.972 & \begin{tabular}[c]{@{}c@{}}-0.001\\ (-0.240, 0.264)\end{tabular} & \begin{tabular}[c]{@{}c@{}}0.30\\ (0.15, 0.41)\end{tabular} & \begin{tabular}[c]{@{}c@{}}0.28\\ (0.03, 1.45)\end{tabular} & \begin{tabular}[c]{@{}c@{}}0.14\ (0.01, 0.73)\end{tabular} & \begin{tabular}[c]{@{}c@{}}0.24\\ (0.10, 0.55)\end{tabular} & \begin{tabular}[c]{@{}c@{}}0.088\\ (0.055, 0.135)\end{tabular} & \begin{tabular}[c]{@{}c@{}}0.60\\ (0.49, 0.74)\end{tabular} & \begin{tabular}[c]{@{}c@{}}0.28\\ (0.15, 0.62)\end{tabular} & \\ \hline
IID-AR1 & 3186 & 3347 & 0.983 & \begin{tabular}[c]{@{}c@{}}-0.006\\ (-0.225, 0.235)\end{tabular} & \begin{tabular}[c]{@{}c@{}}0.30\\ (0.17, 0.42)\end{tabular} & \begin{tabular}[c]{@{}c@{}}0.59\\ (0.44, 0.78)\end{tabular} & \begin{tabular}[c]{@{}c@{}}0.068\\ (0.009, 0.233)\end{tabular} & \begin{tabular}[c]{@{}c@{}}0.24\\ (0.10, 0.57)\end{tabular} & \begin{tabular}[c]{@{}c@{}}0.085\\ (0.051, 0.136)\end{tabular} & \begin{tabular}[c]{@{}c@{}}0.43\\ (0.36, 0.51)\end{tabular} & \begin{tabular}[c]{@{}c@{}}0.26\\ (0.14, 0.56)\end{tabular} & 
\begin{tabular}[c]{@{}c@{}}0.88\\ (0.81, 0.92)\end{tabular} \\ \hline
ICAR-AR1 & \textbf{3154} & \textbf{3302} & \textbf{0.967} & \begin{tabular}[c]{@{}c@{}}-0.007\\ (-0.208, 0.220)\end{tabular} & \begin{tabular}[c]{@{}c@{}}0.32\\ (0.18, 0.43)\end{tabular} & \begin{tabular}[c]{@{}c@{}}0.51\\ (0.34, 0.71\end{tabular} & \begin{tabular}[c]{@{}c@{}}0.072\\ (0.010, 0.235)\end{tabular} & \begin{tabular}[c]{@{}c@{}}0.23\\ (0.10, 0.54)\end{tabular} & \begin{tabular}[c]{@{}c@{}}0.089\\ (0.056, 0.136)\end{tabular} & \begin{tabular}[c]{@{}c@{}}0.51\\ (0.41, 0.69)\end{tabular} & \begin{tabular}[c]{@{}c@{}}0.26\\ (0.14, 0.54)\end{tabular} & 
\begin{tabular}[c]{@{}c@{}}0.92\\ (0.86, 0.96)\end{tabular}\\ \hline
\end{tabular}
}
\end{adjustwidth}

\bigskip \bigskip

\centering
\caption{Proportions of variance explained by each random effect component (percent). } 
\label{Chap3TabVarSumm}
\begin{tabular}{lcccccc}
  \hline
\multirow{2}{*}{Model} & ICAR & Space IID & RW2 & Time IID & Space$\times$Time & Survey \\ 
 & $\sigma_{\alpha}^2$ & $\sigma_{\gamma}^2$ & $\sigma_{\delta}^2$ & $\sigma_{\tau}^2$ & $\sigma_{\phi}^2$ & $\sigma_{\epsilon}^2$ \\
  \hline
IID-IID & 80.8 & 1.6 & 3.8 & 0.4 & 8.7 & 4.7 \\ 
  ICAR-IID & 79.4 & 1.7 & 3.7 & 0.6 & 9.3 & 5.2 \\ 
  IID-RW2 & 12.5 & 1.4 & 3.4 & 0.8 & 75.6 & 6.3 \\ 
  ICAR-RW2 & 17.7 & 4.2 & 2.3 & 0.3 & 72.0 & 3.5 \\ 
  IID-AR1 & 72.5 & 1.4 & 4.5 & 0.6 & 15.2 & 5.8 \\ 
  ICAR-AR1 & 52.7 & 1.9 & 4.7 & 0.8 & 33.1 & 6.9 \\ 
   \hline
\end{tabular}

\end{sidewaystable}

Table \ref{Chap3TabVarSumm} contains the proportions of variance explained by each of the random effects in the space-time smoothing model. The ``ICAR'' column corresponds to the structured spatial effects; ``Space IID'' to the unstructured spatial effects; ``RW2'' to the structured temporal trends; ``Time IID'' to the unstructured temporal effects; and ``Space$\times$Time'' to the respective space-time interaction terms in each model. Across all models, the combined effect from the RW2 and Time IID terms accounts for an average of 4.3\% of the variability, indicating that the overall trend in the RI-specific MCV1 coverage remains relatively stable during the time period of interest. The survey-specific effects and the spatial IID effects also account for relatively small proportions of the variability. The majority of the variation is explained by the structured spatial ICAR terms and the space-time interactions: the combined effect of the two terms accounts for an average of 88.5\% of the variability, and is relatively consistent across all models. The relative proportions explained by the space-time interaction term depend on the specification of the interaction structure. This suggests that the distribution of variability between the spatial ICAR term and the space-time interaction can be sensitive to model specification. It also implies that spatial variation plays an important role in explaining the RI-specific MCV1 coverage. 

\begin{figure}
	\centering
	 \includegraphics[width=0.4\textwidth]{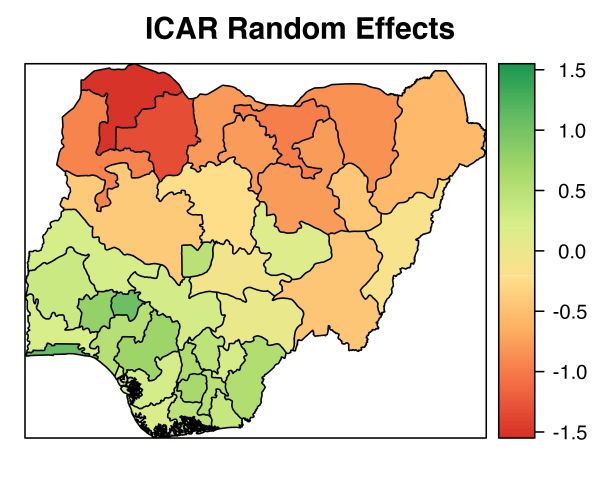}
	 \includegraphics[width=0.4\textwidth]{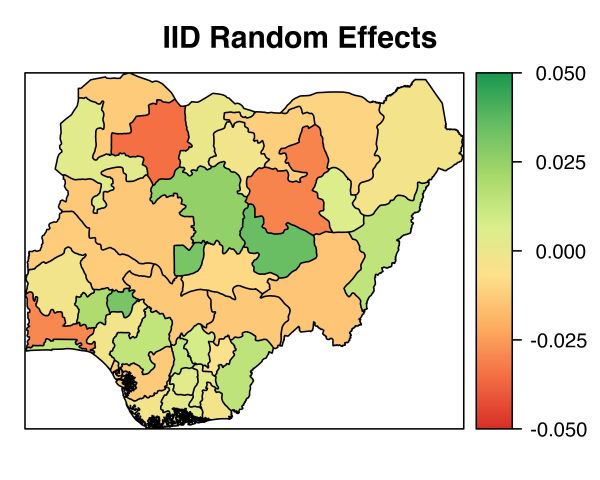}
	\caption{The posterior medians of the ICAR spatial random effects $\alpha_i$ (left) and the IID spatial random effects $\gamma_i$ (right) from the fitted \textit{ICAR-AR1 Model}. }
	\label{Chap3FigSpat}
\end{figure}

We now focus on the \textit{ICAR-AR1 Model} and visualize the estimated random effects in the model. The posterior medians of the ICAR spatial random effects $\alpha_i$ and the IID spatial random effects $\gamma_i$ are shown in Figure \ref{Chap3FigSpat}. The ICAR random effects show apparent spatial dependencies across states, with the northern states having lower estimated values than the southern states. In contrast, and as expected, the IID random effects show no apparent spatial pattern, and their magnitudes are much smaller compared to the ICAR random effects. This suggests that the overall spatial variation in MCV1 coverage is mostly structured, agreeing with the breakdown of variation discussed above.

\begin{figure}
	\centering
	\includegraphics[width=0.45\textwidth]{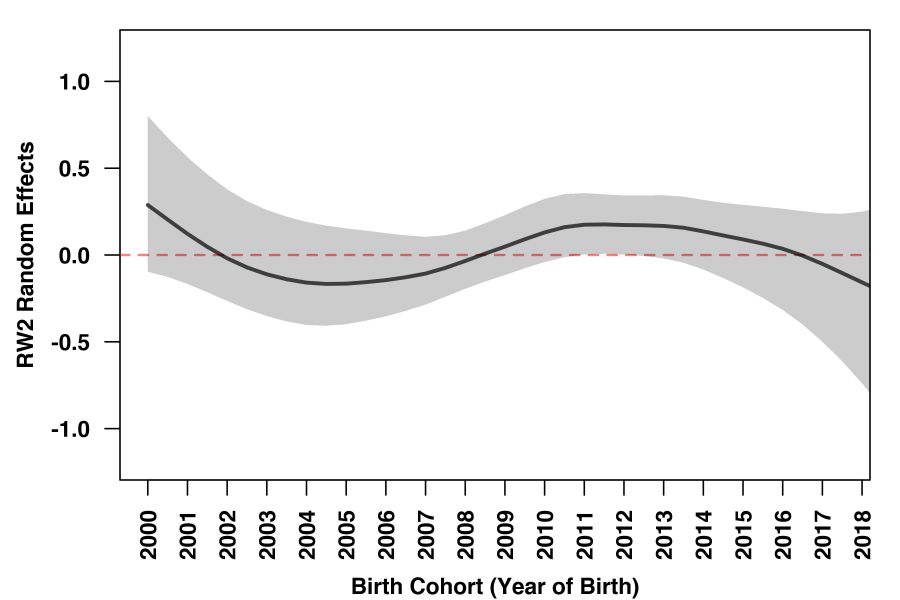}
	\includegraphics[width=0.45\textwidth]{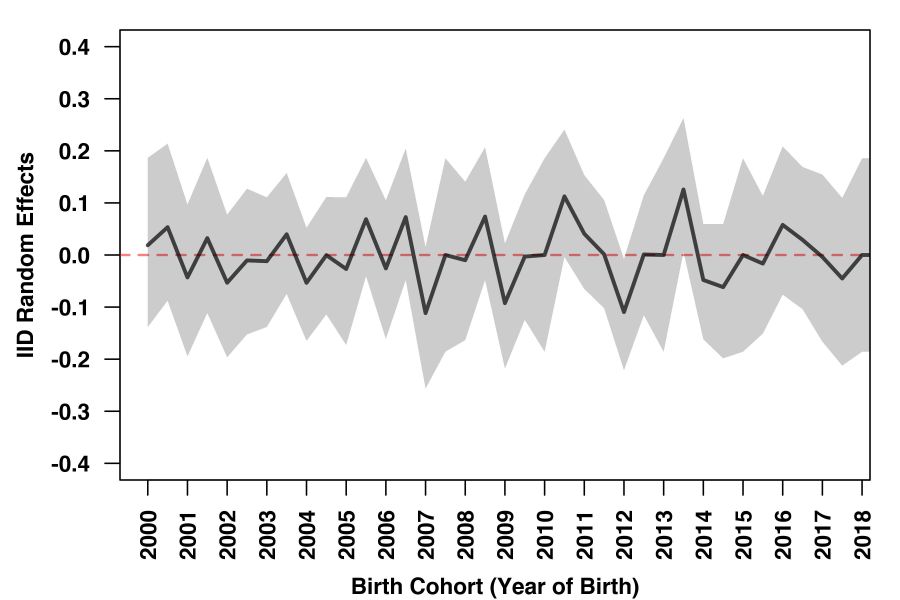}
	\caption{The posterior medians and 95\% CIs of the RW2 temporal random effects $\delta_b$ (left) and the IID temporal random effects $\tau_b$ (right) from the fitted \textit{ICAR-AR1 Model}. }
	\label{Chap3FigTemp}
\end{figure}

The posterior medians of the RW2 temporal random effects $\delta_b$ and the IID random effects $\tau_b$ are shown in Figure \ref{Chap3FigTemp}, along with the 95\% point-wise CI envelopes. The RW2 temporal effects show slight downward trends at the beginning and the end of the time period, and a slight upward trend in the middle. Due to data sparsity, the uncertainties associated with the estimates at the beginning and the end of the time period are relatively large compared to those in the middle of the time period. 

\begin{figure}
	\centering
	 \includegraphics[width=0.27\textwidth]{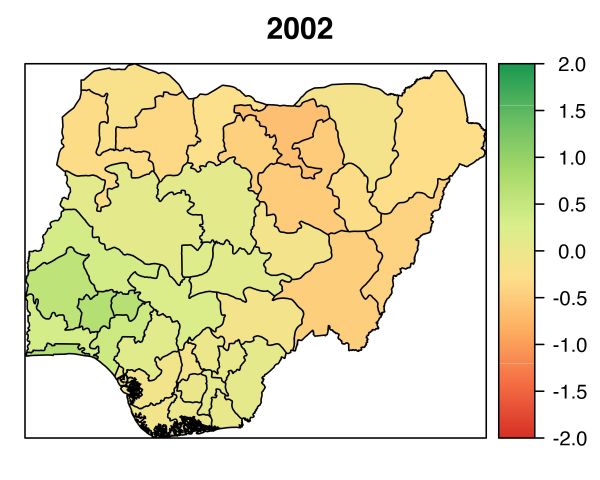}
	 \includegraphics[width=0.27\textwidth]{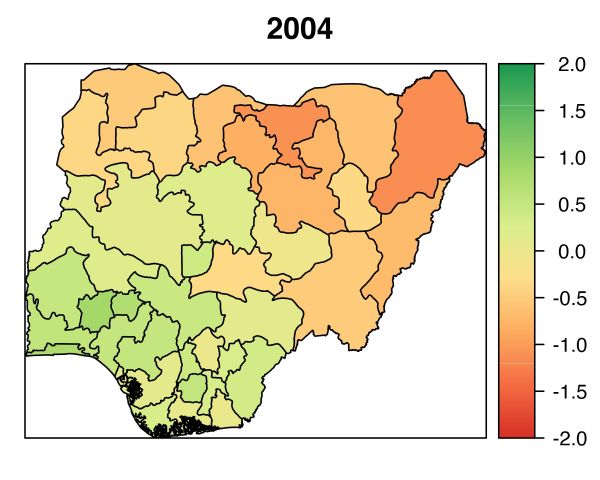}
	 \includegraphics[width=0.27\textwidth]{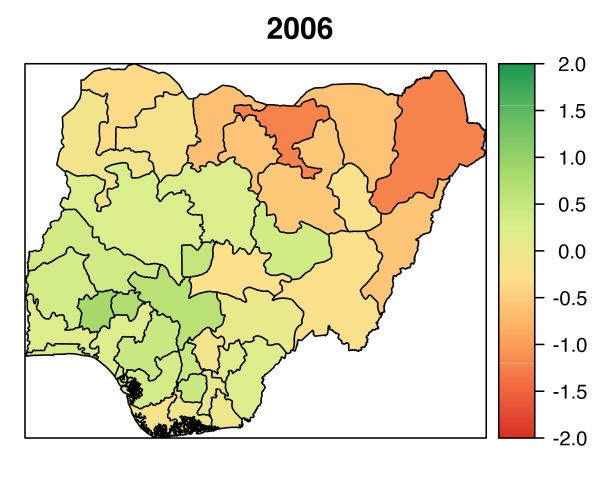}
	 \includegraphics[width=0.27\textwidth]{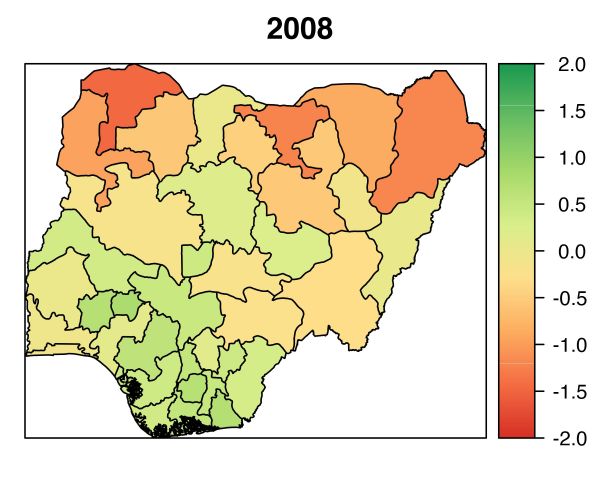}
	 \includegraphics[width=0.27\textwidth]{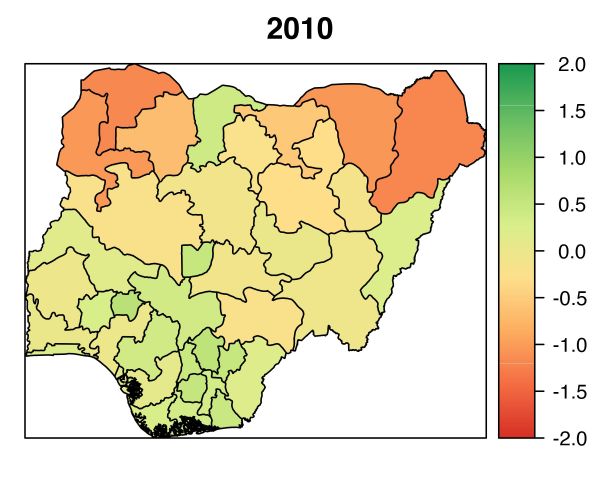}
	 \includegraphics[width=0.27\textwidth]{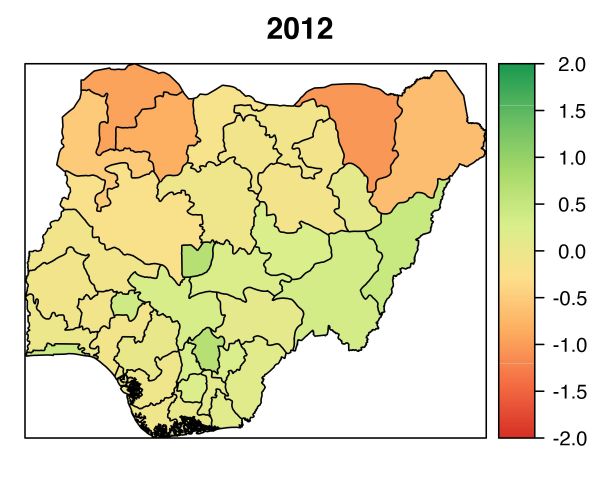}
	 \includegraphics[width=0.27\textwidth]{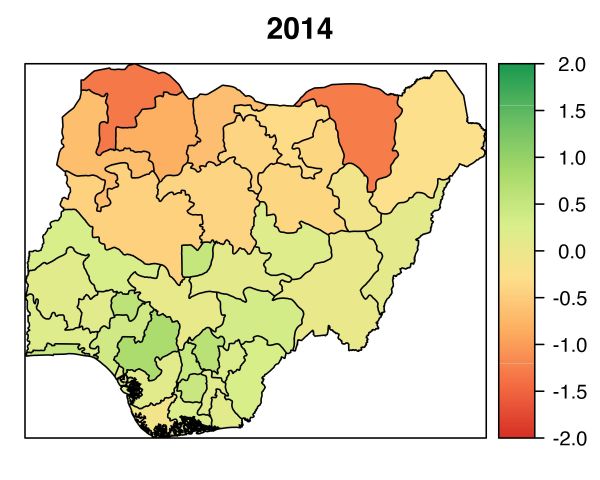}
	 \includegraphics[width=0.27\textwidth]{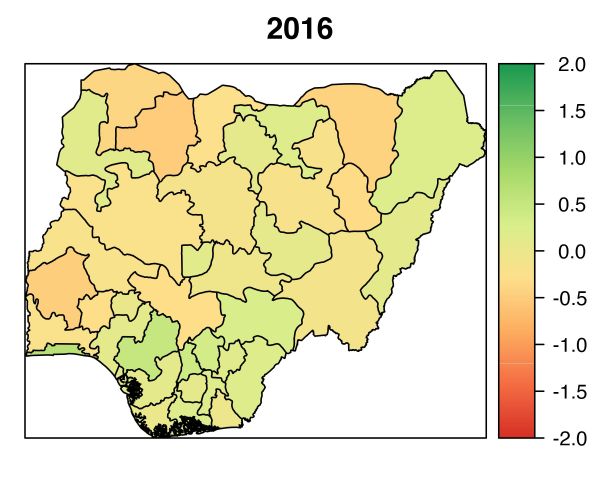}
	 \includegraphics[width=0.27\textwidth]{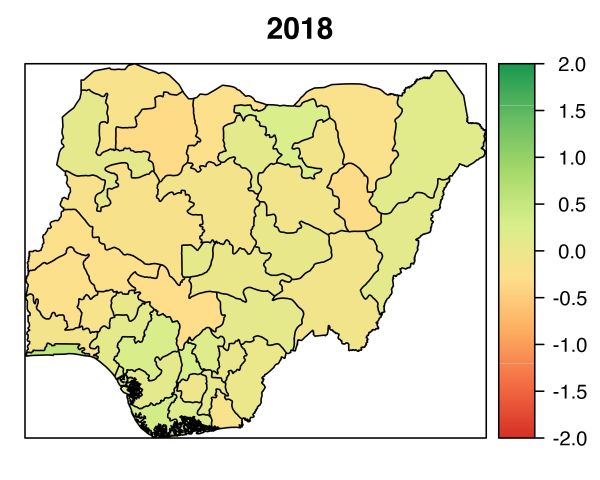}
	\caption{The posterior medians of the space-time interactions  $\phi_{ib}$ across space over selected birth cohorts from the fitted \textit{ICAR-AR1 Model}. Specifically, each map shows the estimated interaction for the cohorts born in the first 6 months of the corresponding year.}
	\label{Chap3FigSTint}
\end{figure}

\begin{figure}
	\centering
	 \includegraphics[width=1\textwidth]{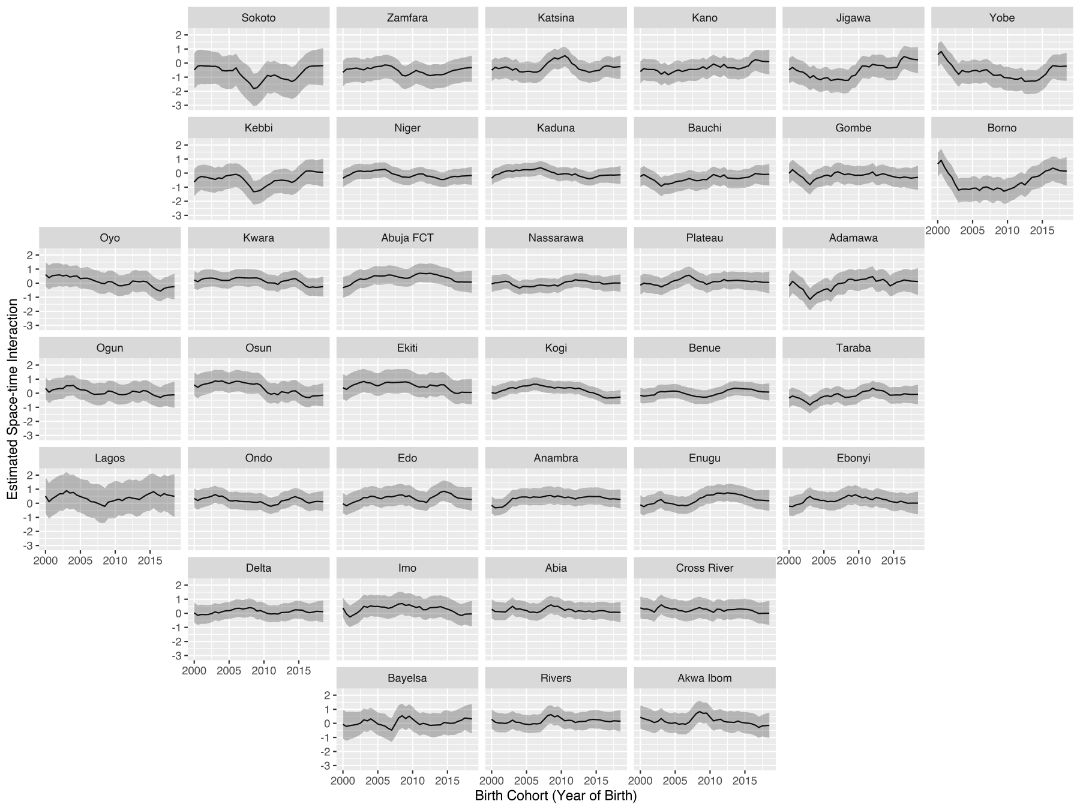}
	\caption{The posterior medians of the space-time interactions $\phi_{ib}$ over time in all states from the fitted \textit{ICAR-AR1 Model}.}
	\label{Chap3FigSTint1}
\end{figure}

In Figure \ref{Chap3FigSTint} and Figure \ref{Chap3FigSTint1}, we show the estimated space-time interactions across space and over time for all states. At each time point, the spatial dependency across states is moderate, with northern states having lower values. The temporal trends in most states were relatively flat without large fluctuations, except for a few northern states in which there were large changes over time. 

\begin{figure}
	\centering
	 \includegraphics[width=0.6\textwidth]{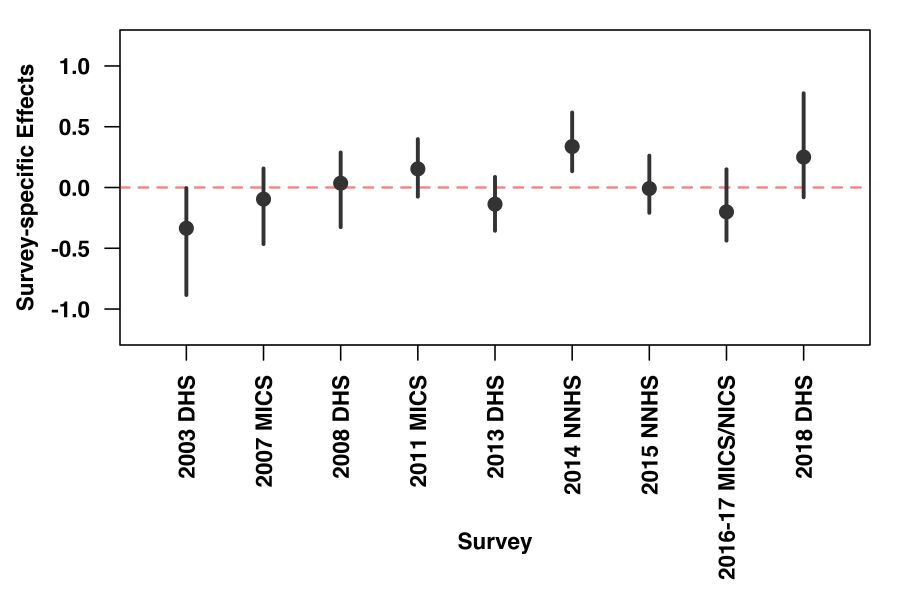}
	\caption{The posterior medians and 95\% CIs of the survey-specific effects $\epsilon_s$. }
	\label{Chap3FigSvy}
\end{figure}

The estimated survey-specific effects $\epsilon_s$ are shown in Figure \ref{Chap3FigSvy}. The estimates suggest that survey effects are not so large --- a result that is also reflected in Table \ref{Chap3TabVarSumm}, where the survey effects are the third largest contributor to the variance explained by the random effects (6.9\%) in the \textit{ICAR-AR1 Model} model. 

\begin{figure}
	\centering
	 \includegraphics[width=1\textwidth]{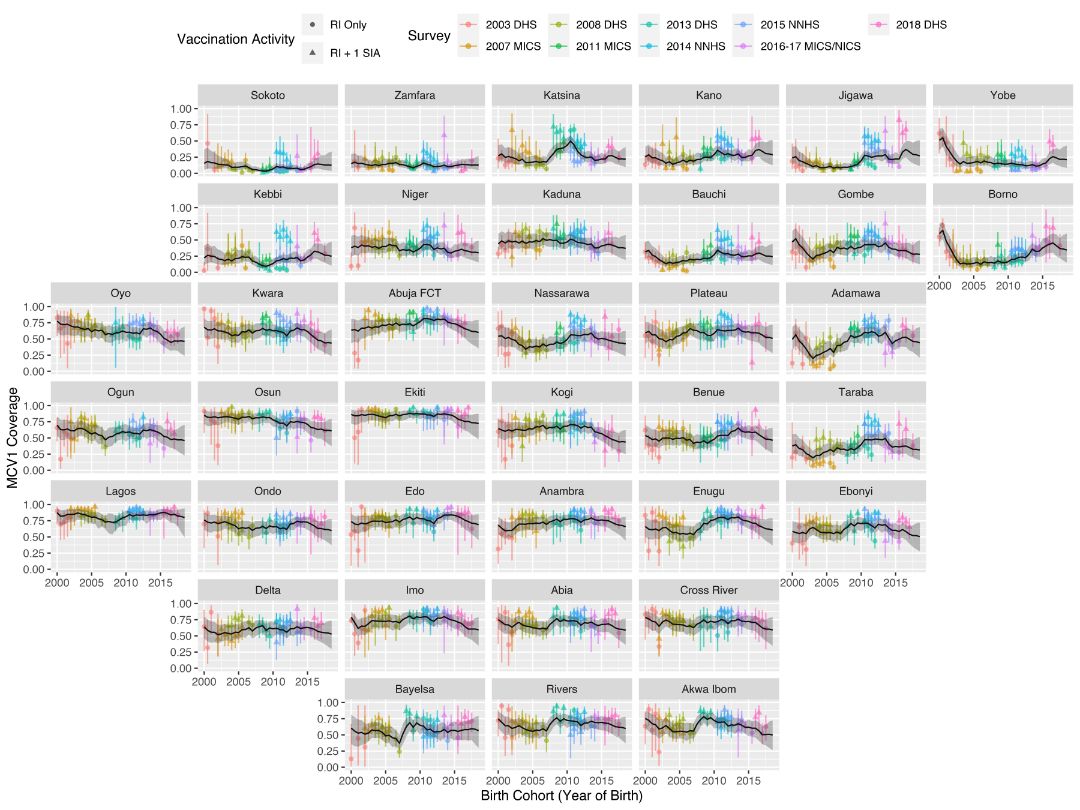}
	\caption{The posterior medians of the state-level RI-specific MCV1 coverage (dark grey lines) and the associated 95\% CIs (light grey ribbons) for the birth cohorts born between 2000 and 2018 based on the fitted \textit{ICAR-AR1 Model}. The design-based overall MCV1 coverage estimates were also shown for references. }
	\label{Chap3FigResults}
\end{figure}

Based on the fitted \textit{ICAR-AR1 Model}, we calculated the state-level RI-specific MCV1 coverage estimates and the associated 95\% CIs for the birth cohorts born between 2000 and 2018 using the procedure described in Section \ref{Chap3Eqn4}. Figure \ref{Chap3FigResults} shows the model results, along with the design-based overall MCV1 coverage estimates from each survey for reference. There is considerable spatial variation in the estimated RI-specific MCV1 coverage, with southern states generally having higher coverage than the northern states. In most states, the RI-specific coverage remained at a constant level without significant changes in trend between 2003 and 2015. The 95\% CIs associated with the RI coverage estimates during this time period are relatively narrow. In contrast, the RI-specific coverage estimates in 2000--2002 and 2016--2018 show greater temporal variation, but they are associated with wider CIs. Therefore, although there seem to be a slight downward trend in many states, one should be mindful of the uncertainties associated with the trend, especially when considering prediction for the 2018 birth cohorts. 

In summary, this analysis illustrates how our model is able to combine information from multiple household surveys to produce spatially- and temporally-smoothed RI-specific MCV1 coverage estimates while accounting for the impact of SIAs as well as survey-specific effects.


\subsection{Sensitivity analysis: using 12-month birth cohorts}

We chose to use 6-month birth cohorts in our model as they gave a good balance between two competing objectives: on one hand, we want larger birth cohorts so that there would be enough samples from each sub-national area in each survey to provide design-based coverage estimates with reasonably high precision; on the other hand, smaller birth cohorts are more desirable because they are more likely to have children with the same RI and SIA history, and hence be eligible for analysis. Here we present a sensitivity analysis using 12-month birth cohorts. 

\begin{figure}
	\centering
	 \includegraphics[width=1\textwidth]{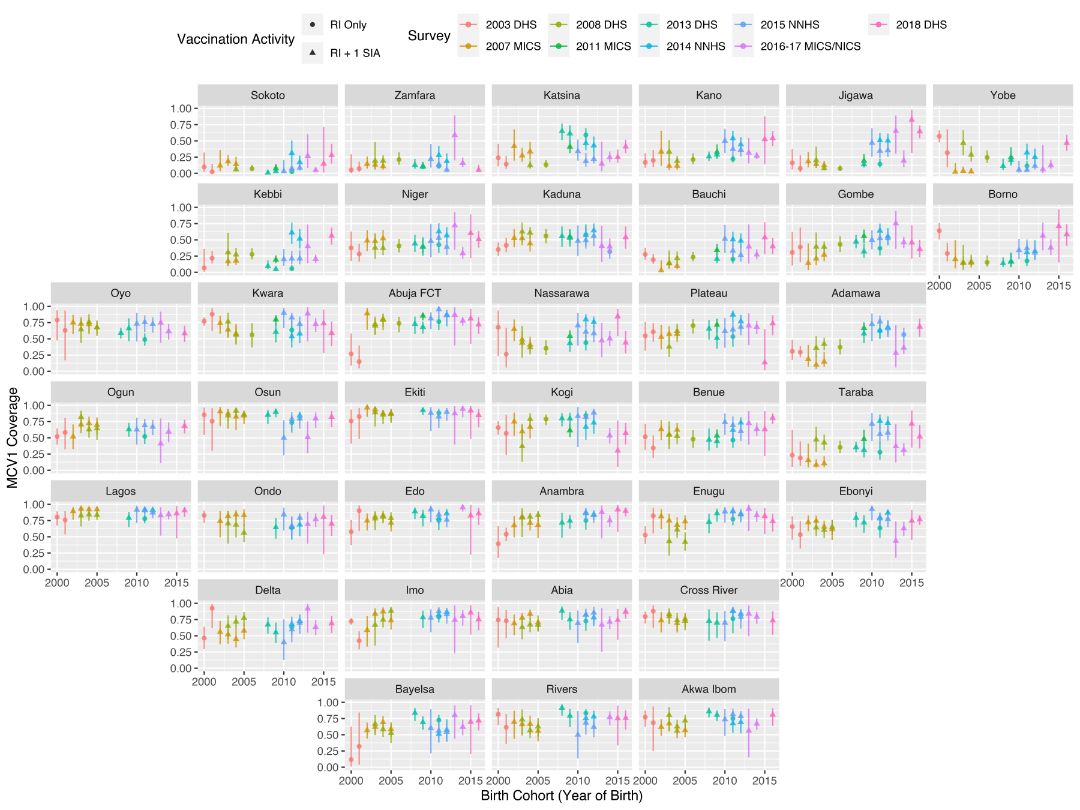}
	\caption{Design-based HT estimates of the overall MCV1 coverage across birth cohorts from each survey for Nigeria's 37 states in the sensitivity analysis with 12-month birth cohort data.}
	\label{Chap3FigDirest_12mSA}
\end{figure}

We discretize time into 12-month intervals and let the beginning of 2000 be the reference starting point. We carry out the same data processing step outlined in Section \ref{Chap3SecDataproc} and calculate the design-based HT estimates of the state-level overall MCV1 coverage for each eligible birth cohort in each survey. Figure \ref{Chap3FigDirest_12mSA} shows the resultant estimates with the associated confidence intervals across birth cohorts for Nigeria's 37 states. As expected, the estimates for 12-month birth cohorts generally have narrower 95\% CIs compared to the estimates for 6-month birth cohorts shown in Figure \ref{Chap3FigDirest}. However, only 61900 of the 86033 (71.9\%) survey samples that were eligible for the previous analysis remain eligible for this analysis, and the proportion of the \textit{RI Only} design-based coverage estimates (among the \textit{RI Only} and \textit{RI + 1 SIA} design-based coverage estimates) decreased from 608/1802 (33.7\%) to 133/760 (17.5\%). In particular, 35 of the 37 states do not have any \textit{RI Only} design-based coverage estimate for any birth cohort born after 2012. As a result, we see that many southern states do not have any eligible data for birth cohorts born between 2006 and 2008, and a majority of the design-based overall MCV1 coverage estimates were influenced by a previous SIA. 

The six space-time smoothing models specified in the main analysis were fitted to the design-based overall MCV1 coverage estimates for the 12-month birth cohorts. The posterior medians and 95\% CIs of the model parameters as well as the model assessment metrics are reported in Table S3 of the Supplementary Material. Compared to the results in the main analysis, the estimates of the intercept $\beta_0$ are considerably higher and the estimates of the coefficient for the SIA indicator $\beta_1$ are lower. The 95\% CIs for these two parameters are also wider across all models. These results are likely due to the fact that less data, especially \textit{RI Only} design-based coverage data, is used in this analysis. Since less information is available for the model to draw inference on the parameters, the resultant estimates differ from those in the main analysis and have higher associated uncertainties. 

Similar to the main analysis, the \textit{ICAR-AR1 Model} performs the best among the six models based on all three model metrics considered. The proportions of variance explained by the various random effects also show similar patterns compare to the main analysis. The details are provided in Table S4 of the Supplementary Material. 

We now focus on the \textit{ICAR-AR1 Model} and compare the results from fitting this model to the 12-month birth cohort data with the \textit{ICAR-AR1 Model} results in the main analysis. The estimated ICAR and IID spatial random effects and the survey specific effects show very similar patterns between the two analyses. The estimated space-time interactions also show similar general trends, but some small-scale fluctuations shown in the main analysis are missing in this analysis, especially during time periods where \textit{RI Only} design-based coverage data are not available. The estimated RW2 temporal random effects in this analysis are following a slightly different trend compared to the corresponding estimates in the main analysis: instead of showing a slight downward trend after 2012, the estimated RW2 random effects follow a slight increasing trend. In addition, the IID temporal random effects have considerably wider uncertainty intervals than those in the main analysis. Plots of these estimates and the associated uncertainties can be found in Section S5 of the Supplementary Material. 

In Figure \ref{Chap3FigResults_6m12mSA}, we compare the state-level RI-specific MCV1 coverage estimates produced by fitting the \textit{ICAR-AR1 Model} to the 6-month and 12-month birth cohort data. The estimates from the 12-month birth cohort analysis are generally higher than those from the 6-month analysis. In particular, the differences between the two analyses tend to be large during time periods in which (1) there is no design-based coverage data (e.g., between 2006 and 2008), or (2) there is no \textit{RI Only} coverage data (e.g., after 2012) for fitting the space-time smoothing model in the 12-month birth cohort analysis. The pattern we see in this comparison, as well as the difference in parameter and random effect estimates we showed earlier, are likely due to the fact that using 12-month birth cohorts resulted in considerably less survey data being eligible for the smoothing model, hence some temporal patterns that were captured in the 6-month birth cohort analysis were unable to be estimated in this analysis. 

\begin{figure}
	\centering
	 \includegraphics[width=1\textwidth]{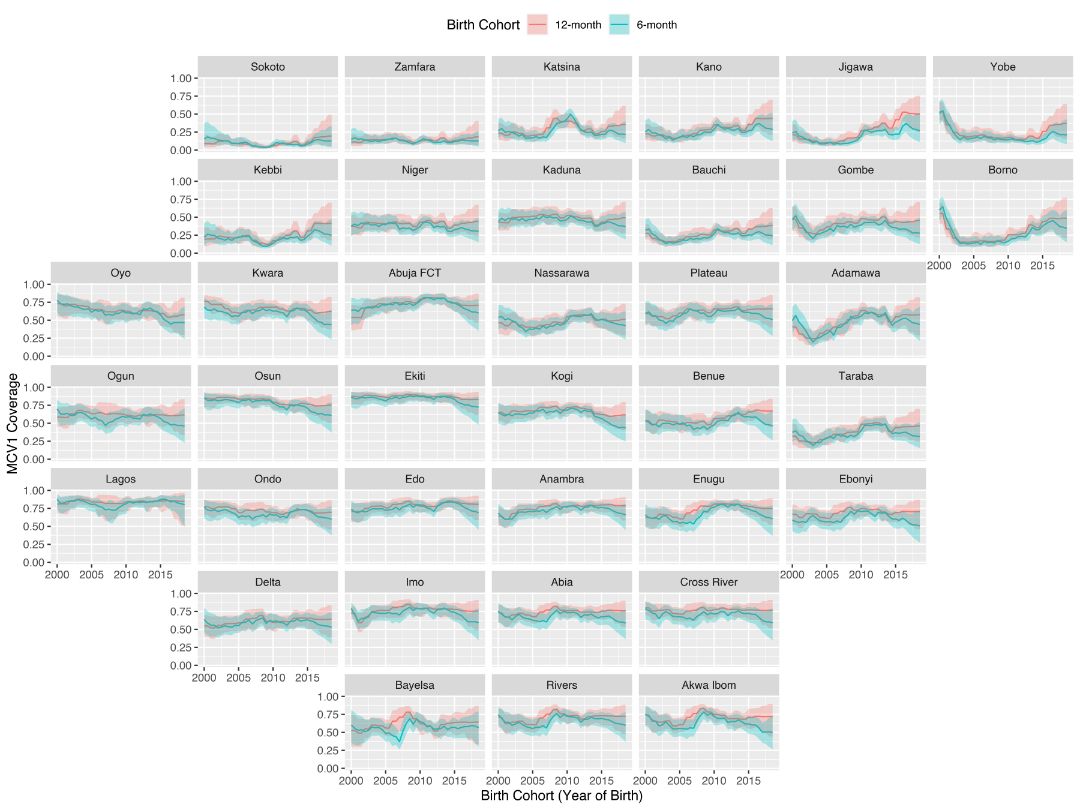}
	\caption{The posterior medians of the state-level RI-specific MCV1 coverage (dark grey lines) and the associated 95\% CIs (light grey ribbons) for the birth cohorts born between 2000 and 2018 based on the fitted \textit{ICAR-AR1 Model}. }
	\label{Chap3FigResults_6m12mSA}
\end{figure}

 
\section{Discussion}  \label{Chap3SecDisc}

In this paper, we developed a space-time smoothing model for estimating RI-specific MCV1 coverage at the sub-national level over time using data from household surveys. The proposed method is able to account for the impact of SIAs on the overall MCV1 coverage measured by cross-sectional surveys, and makes efficient use of data from not only birth cohorts who only had RI opportunities, but also those who had RI plus one SIA opportunity at the time of surveys. In addition, our model can be applied to data from multiple surveys with different sampling designs and construct coverage estimates with uncertainty that reflects the various data collection schemes. Any survey-specific effects such as sampling biases are also accounted for when predicting underlying RI coverage over time. We applied our method to analyze data from nine household surveys conducted in Nigeria and constructed the RI-specific MCV1 coverage in each of Nigeria's 37 states from 2003 to 2018. 

Since the DHS and MICS regularly conduct household surveys in most LMICs, our method is readily applicable to many countries with high measles burden to provide reliable RI-specific MCV1 coverage estimates over time to track routine vaccine delivery system strength. It helps identify areas in which the RI program is the weakest and facilitates balanced implementation of RI and SIAs to improve overall measles vaccination coverage. In addition, our method can be easily modified to produce sub-national coverage estimates for other types of vaccines, including those not influenced by SIAs such as the diphtheria, pertussis, and tetanus (DPT) vaccines and the Bacille Calmette-Guerin (BCG) vaccine. 

Under our framework, the RI-specific MCV1 coverage has a clear and sensible definition --- the percentage of children in a birth cohort who receive a first MCV dose according to the RI schedule. This definition is particularly relevant in the context of building measles epidemiological models and evaluating SIA efficacy, where the RI-specific coverage is an essential input for calculating the number of children entering the susceptible pool over time \citep{verguet2015controlling, thakkar2019decreasing, dong2020estimating}. We emphasize here that the denominator of this coverage only includes children who are still alive at their scheduled RI date, i.e., at the age of 9 months. In other words, our coverage definition excludes those who died before they become eligible to receive MCV1 through RI, and hence, does not require consideration of the mortality rate among those under the age of 9 months. 

Our method does make assumptions about the mortality rates among children above the age of 9 months. Specifically, we assume that there is negligible difference in mortality rate between the children who are vaccinated against measles and the children who are not. This allows us to assume the underlying MCV1 coverage within each birth cohort remains constant as the children grow older, unless they have an SIA opportunity. As such, we can treat each cross-sectional survey as a snap-shot measurement of the same underlying MCV1 coverage within a birth cohort at the time of survey interview, and use survey data collected from children in older age groups for our model. However, if the unvaccinated children suffer significantly higher mortality than vaccinated children, the \textit{proportion} of the surviving children in a birth cohort who are vaccinated would increase as the children grow older. In this case, the underlying MCV1 coverage within each birth cohort no long remains unchanged over time, and the coverage estimates from surveys implemented at different times among the same birth cohort are measuring different underlying coverage values. Therefore, a more complex model as well as additional mortality data are required to properly account for the impact of discrepancies in mortality rate between vaccinated and unvaccinated children, which would be an important topic for future research. 

In addition, our model also assumes that children have their RI opportunity according to the stated schedule, and SIAs were carried out promptly as recorded in the calendar. These assumptions may not hold in reality, especially in settings where vaccine stock outs or disruption of services lead to delayed RI and SIAs. Their impact on RI-specific coverage estimation would depend on when and where SIAs and surveys were carried out. Further simulation studies need to be conducted to investigate the issue under various scenarios. More recently, the COVID-19 pandemic has disrupted routine immunization services on an unprecedented scale, and many countries have temporarily suspended measles SIAs due to risk of transmission and the need to maintain physical distancing \citep{who_covid}. Region-specific adjustment will be required to account for the impact of the pandemic when estimating measles vaccination coverage during the relevant time periods. 

Last, but not least, we assume there is negligible systematic recall bias across age groups in surveys. However, it is possible that care givers would under- or over-report doses for older children due to memory recall. Previous studies gave mixed results regarding the presence, direction and magnitude of care giver recall bias in vaccination coverage surveys \citep{porth2019childhood, valadez1992maternal, ramakrishnan1999magnitude, hu2019validity, binyaruka2018validity}, and the results vary greatly by geographical regions. Future research could focus on how to detect and account for recall bias, especially given the specific context of measles vaccination.




\begin{supplement}
\stitle{Supplement to ``Space-time smoothing models for sub-national measles routine immunization coverage estimation with complex survey data''}
\slink[doi]{TBD}
\sdatatype{.pdf}
\sdescription{}
\end{supplement}


\bibliographystyle{imsart-nameyear}
\bibliography{References} 

\end{document}